\documentclass[sigconf, authorversion]{acmart} 

\usepackage{url}
\usepackage{graphicx}
\usepackage{booktabs}
\usepackage{tabularx}
\usepackage{booktabs}
\usepackage{balance}

\copyrightyear{2024}
\acmYear{2024}
\setcopyright{rightsretained}
\acmConference[ARES 2024]{The 19th International Conference on Availability, Reliability and Security}{July 30-August 2, 2024}{Vienna, Austria}
\acmBooktitle{The 19th International Conference on Availability, Reliability and Security (ARES 2024), July 30-August 2, 2024, Vienna, Austria}\acmDOI{10.1145/3664476.3670456}
\acmISBN{979-8-4007-1718-5/24/07}

\begin{document}

\title{An Exploratory Case Study on Data Breach Journalism}

\author{Jukka Ruohonen}
\affiliation{\institution{University of Southern Denmark}
  \city{S{\o}nderborg}\country{Denmark}}
\email{juk@mmmi.sdu.dk}
\author{Kalle Hjerppe}
\affiliation{\institution{University of Turku}
  \city{Turku}\country{Finland}}
\email{kphjer@utu.fi}
\author{Maximilian von Zastrow}
\affiliation{\institution{University of Southern Denmark}
  \city{S{\o}nderborg}\country{Denmark}}
\email{maxvz@mmmi.sdu.dk}

\begin{abstract}
This paper explores the novel topic of data breach journalism and data breach
news through the case of \textit{databreaches.net}, a news outlet dedicated to
data breaches and related cyber crime. Motivated by the issues in traditional
crime news and crime journalism, the case is explored by the means of text
mining. According to the results, the outlet has kept a steady publishing pace,
mainly focusing on plain and short reporting but with generally high-quality
source material for the news articles. Despite these characteristics, the news
articles exhibit fairly strong sentiments, which is partially expected due to
the presence of emotionally laden crime and the long history of sensationalism
in crime news. The news site has also covered the full scope of data breaches,
although many of these are fairly traditional, exposing personal identifiers and
financial details of the victims. Also hospitals and the healthcare sector stand
out. With these results, the paper advances the study of data breaches by
considering these from the perspective of media and journalism.
\end{abstract}

%



\maketitle

\section{Introduction}

Data breaches are a constant and enduring foul hindrance to consumers, citizens,
companies, public sector organizations, and others in the modern digitalized
societies. With these cyber attacks, cyber criminals steal sensitive personal
data of consumers and citizens, valuable business data of companies, and data
from various other organizations. Despite continuous investments to cyber
security, data breaches have in fact again increased in recent
years~\cite{Madnick24}. Although data breaches are a technical research topic,
these have frequently been investigated also from other
perspectives. Organizational aspects have been a common research topic together
with human security perspectives. Societal perspectives have been rarer.

Data breaches are often, but not always, about cyber crime, and crime is very
much also a societal topic. Crime and media, in turn, are closely associated,
and media too is very much a societal topic. These assertions provide a
rationale for the present paper to examine the previously more or less
unexplored topic of data breach journalism. The motivation is built around the
long and debatable history of traditional (non-cyber) crime news and crime
journalism. The frequent sensationalism and negativity in this type of news and
journalism provide a motivation to not only examine the data breaches reported
but also the style of this reporting. In addition, the discussion in the opening
Section~\ref{sec: background} addresses the potential consequences from data
breach journalism, including for the victims of data breaches, media and
journalists, and a society at large. Subsequently, Section~\ref{sec: data and
  methods} outlines the materials and methods for the case study. The
exploratory text mining results are presented in Section~\ref{sec: results}. The
final Section~\ref{sec: discussion} summarizes the key results, enumerates a few
limitations, and outlines paths for further research.

\section{Background and Framing}\label{sec: background}

\subsection{Crime News and Crime Journalism}\label{subsec: crime news}

Data breaches can occur in multiple ways. These include also factors related to
human behavior and organizational postures~\citep{Schlackl22}. Both malicious
motives and innocent mistakes may be involved. An accountant may accidentally
disclose sensitive financial data, a salesperson may unnecessarily carry
sensitive customer data that leaks, management may be unaware of the security
risks in the face of mergers and acquisitions, a disgruntled and indebted
employee may sell sensitive data to an adversary, and so forth and so on.

More often, however, data breaches nowadays occur online through exploitation of
technical vulnerabilities in software systems. Such exploitation shifts the
focus more explicitly to cyber crime. This characterization is true also in
media within which big global data breaches frequently attain flashy
headlines. Furthermore, data breaches form their own specific cyber security
news niche already because these are almost always tied to specific events,
unlike many other cyber threat narratives in media, such as those related to
national security and espionage~\cite{Boholm21}. In other words, a data breach
in itself is almost always the specific event to which the given data breach
news build upon. A data breach event typically also remains as the focal point
in follow-up news.

Crime news and crime journalism have a long history, as does the fields that
study this form of journalism. Already in the early 20th century (if not
earlier), it was common to speculate about and argue over the role crime plays
and should play in media and journalism, how crime should be reported, what
function crime news fulfills or should fulfill in a society, and what
consequences it may entail, among other related topics of interest. While crime
journalism was an integral historical part in the popularization of media, it
was also commonly seen as a hindrance to raising the status of journalism and
the professionalization of journalists along the way~\cite{Rossland07}. To this
end, it was commonly argued that crime news should move away from
sensationalism, which may increase the risk of emulations and copycats, among
other things, toward constructive reporting that informs the people about the
dangers in their neighborhoods, possibly leading to beneficial regulatory or
other changes for both the public and law enforcement~\cite{Rice29}. This
characterization captures many typical tenets of crime journalism even today:
crime is newsworthy because it is, by construction, sensational, extraordinary,
and disruptive; crime tells also about the political realities people live
under; and crime news often require resonance with the everyday life of people,
and a journalistic way to achieve such resonance is to rely on geographical or
historical proximity~\cite{Higgins22}. It is not difficult to relate these
tenets to cyber crime and data breaches in particular. Data breaches are often
sensational because they showcase the banality of the cyber (in)security of many
organizations; data breaches can act---and have acted---as a major political
force pushing new cyber security regulations onward; and historical proximity is
guaranteed because everyone in practice knows a previous breach---if not a
victim of such a~breach.

Recently, the debate has also touched victims of crime and victimization,
meaning the process of becoming a victim or being victimized by someone. Here,
to begin with, exposure to sensational (or negative) crime news tends to
increase the perception of people about the probability of becoming a victim of
a crime~\cite{Velasquez20}. This effect is presumably partially due to the
pervasive notion of risk in crime news, which often report on serious and
seemingly random crimes committed by strangers~\cite{OLeary21}. Again, it is not
difficult to draw an analogy to data breaches; if it can happen to everyone, it
can happen to me as well. Though, it must be added that considerable skepticism
has been expressed over whether people consume crime news because they want to
learn about crime statistics and related robust details about crime or because
they want to entertain themselves, protect collective moral identity, moralize
on political conflicts, or something along these lines~\citep{Katz87}. Such
consumer appetite is a central factor behind a common criticism that crime news
tend to create an oversimplified or outright false picture of crime, rather
promoting existing fears, stereotypes, biases, and prejudices~\cite{OLeary21,
  Smolej08}. That said, there is also criticism of the criticism, particularly
with respect to the assumption about the effect of crime news upon
audiences~\cite{Doyle06}. In any case, the use of victims (instead of criminals
or suspects) in crime news may perhaps help to mitigate unintended
consequences. Regarding data breaches, it is indeed not difficult to find
examples of human interest stories whereby a breach is approached through its
consequences to the victims.

Also other affected people may appear in crime news, including, of course, law
enforcement or other public authorities, but also eyewitnesses and people
related to a suspect or a victim often appear in crime news. For instance,
eyewitnesses can be used in crime news to give a mediated witness experience
whereby the consumers of the news align their emotional response to those
affected by a crime~\cite{vanKrieken22}. Although such journalistic techniques
may in theory appear also in cyber crime and data breach news, it is more common
to see a voice from law enforcement in these news. In fact, for reasons soon to
be discussed, data breach news often rely on official statements from law
enforcement as the building blocks for the news. This sourcing may lead to a
known bias, given that a law enforcement perspective is not alone the only
relevant viewpoint to crime~\citep{Sacco95}. Therefore, it is worth further
remarking that not only interviews of law enforcement officials but also
interviews of data protection authorities and cyber security experts sometimes
appear in cyber crime and data breach news.

Finally, it should be emphasized that crime journalism is culturally dependent
with different geographic regions having different traditions and frameworks for
crime news reporting. What works in the United States or Southeast Asia may not
work (or even be legal) in Germany. Also the levels of professionalization likely differ. For instance, in some Nordic countries crime
journalism enjoys the same level of professionalization as other types of
journalism, being also subject to the same ethical and related guidelines
monitored by self-regulatory councils, although, paradoxically, it was also the
criticism on the sensational crime news that partially led to the
professionalization and associated media practices~\cite{Rossland07}. That said,
crime news have increased in volume also in some Nordic countries, and much of
this increase is attributable to tabloids who may be less strict about ethics
and professional practices~\cite{Smolej08}. This point serves to underline that
the long-lasting debate on crime news and crime journalism is still alive today
in most countries.

\subsection{Potential Consequences}\label{subsec: consequences}

In light of the previous discussion, it is helpful to consider the case in a
larger context of data breach journalism and its potential individual, economic,
societal, and other consequences. Thus, the following seven points briefly
consider the potential consequences from data breach journalism in a larger
context.

First, it may be that data breach journalism shares the same sins as crime news
generally. In particular, it may be biased in various ways. For instance, it may
exaggerate the security risks from data breaches, potentially leading to
uncalled fears among the public. Consequently, data breach journalism may
contribute to painting of the Internet as an excessively insecure and dangerous
place. These points align with an argument that cyber crime tends to be often
associated with a moral panic in media, which may also have unintended
consequences, such as inefficient resource allocation or poorly justified legal
powers granted to law enforcement agencies~\cite{Lavorgna19}. However, it is
impossible to evaluate how realistic these points are in practice; the evidence
is in the eye of the beholder.

Second, there is a potential risk of data breach journalism cultivating new
cyber crime through inspiring aspiring cyber criminals, hooligans, and other
perpetrators. Fame is supposedly of high value in cyber crime circles, and big
data breaches are a sure way to attain fame. However---and unlike with some
other types of crime, information technology media generally lacks a true crime
genre, which at least implicitly may glorify crime and criminals, potentially
leading to the copycat behavior, as has been the case with respect to violent
drug crime in some countries~\cite{Rios20}. While there are some media that come
close, such as \textit{Darknet Diaries}, these are rare and supposedly not that
popular among a wider audience. To perhaps exclude high-profile crime
syndicates, in cyber crime the dynamics of fame supposedly largely happen among
cyber criminals themselves in darknet and closed underground forums.

Third, data breach journalism may cause additional harm to the victims. Here, it
is important to stress that many---if not most---data breaches involve two types
of victims: the companies or other organizations who were breached and the
persons, customers, business associates, or other types of stakeholders whose
data was breached. The potential harms differ between the two types. Although
the evidence is not entirely conclusive, data breaches may even affect the stock
prices of the victim companies who were breached~\cite{Spanos16}. Analogously,
losing sensitive data of business associates to criminals is hardly good
business. Regulations such as the General Data Protection Regulation (GDPR) have
also introduced potentially large financial sanctions from data
breaches. Therefore, it is no wonder that data breaches are a significant
reputation threat to companies~\cite{Syed19}. To protect their images, companies
and other organizations typically use different crisis communication strategies
when dealing with data breaches, and these may also affect the narratives
presented in media~\cite{Ruohonen24PRR}. However, it is difficult or even
impossible to say whether a company is still better off with a data breach
compared to a ransomware attack whereby the company's data is encrypted and kept
as a ransom. Nevertheless, in both cases the potential consequences provide a
clear incentive for companies to keep the information disclosed to media as
limited as possible. In terms of the other victim type, a good albeit morbid
case would be the Finnish \textit{Vastaamo} breach. It involved a breach of a
psychotherapy center whose patient records were leaked to the darknet and
further used as extortion material. Given the sensitivity of the data involved
and the psychological state of the victims, the constant news coverage of the
breach may have caused additional stress to the victims---or patients. This case
aligns with arguments on the elevated victimization of data breach victims after
a notification of a breach~\cite{Gibson23}. The example also demonstrates that
data breach journalism involves its own ethical considerations.

Fourth, data breach journalism may also help the user-style victims by raising
awareness among them. Indeed, information about data breaches has been observed
to correlate with security awareness~\citep{Chua21, Mayer23}. A correlation has
been further found to awareness of data protection regulations, including the
GDPR~\cite{YanChua20}. Then: if social security numbers and bank account details
have been breached, a risk of identity thefts may be significant enough to
warrant reporting also in media. By becoming aware of a breach, the persons
affected may then take necessary precautions. Unfortunately, however, this line
of reasoning does not necessarily hold that well because existing results also
hint about poor practical security posture among users even after they have
become aware of an incident through media or by other means~\cite{Bhagavatula21,
  Mayer23}. Another way to look at the awareness question is to focus on the
other victims; data breach journalism may help users also by shedding light
on companies or organizations with haphazard security practices. The major
\textit{Equifax} breach in the United States serves as an example along these
lines, which again resemble the constructive crime journalism ideals.

Fifth, a potential downside of awareness raising is so-called data breach
fatique. Indeed, some studies indicate that at least a portion of audiences has
become tired of constant bombardment of data breach news~\cite{Reitberger17}. To
counter this fatique problem, media outlets might concentrate on bigger data
breaches with feature journalism instead of plain reporting on daily breaches.

Sixth, data breach investigations require trustworthy frameworks and secure
communication channels between the parties involved, including, but not
necessarily limited to, the victim organization, the investigators, and possibly
law enforcement or other governmental agencies~\cite{Thomas22}. If sensitive
details leak to media during an investigation, wild headlines may follow but
these may hamper the given investigation. Such scenarios again underline that
data breach journalism should follow its own ethical practices. Ethics apply
also to the relatively frequent cases in which a data breach is found by a
non-malicious third-party who may report it both to the given organization and
media~\citep{IBM23, Ruohonen24PRR}. In terms of ethics, the issue is rather
similar to voluntary reporting of vulnerabilities to vendors. Moreover,
regulations such as the GDPR have introduced notifications to the natural
persons affected; therefore, leaks to media are likely to occur in any
case. That said, leaks should not downplay ethics in journalism and associated
professional practices.

Last but not least, it is possible to remain true to crime journalism and
maintain that ``news does more good than harm'' \cite{Rice29}. If nothing else,
data breach journalism, especially in the context of the case studied, helps to
keep public records straight. While vigilantism is a potential threat, this line
of argumentation aligns with observations that the public, after having become
aware a cyber crime, may help at resolving the crime, especially when done in
cooperation with law enforcement agencies~\cite{Huey12}. Finally, as there is no
robust global database on data breaches, a systematic journalistic reporting on
breaches can also create rich archival data for data mining.

\subsection{Related Work}\label{subsec: related work}

According to a fairly comprehensive literature search, no directly comparable
previous works exist. Having said that, there are some studies that come close
to the present paper's theme and its framing. Mimicking existing work for
quantifying general crime news~\cite{Lindgren08}, attempts have been made to
profile different cyber crimes through online news
coverage~\citep{Christian22}. According to this data mining branch of research,
moreover, data breaches are often reported negatively in media, including with
respect to the victim organizations~\citep{Mazrouei22, Sinanaj14}. Such results
are hardly surprising in the sense that negative news commonly surpass positive
ones, partially because negativity often drives consumption~\cite{Robertson23,
  Trussler14}. Perhaps due to this demand-side, even many journalism textbooks
tend to rationalize the commonplace reporting of crime, conflict, and bad
news~\cite{Parks19}. These results reinforce the earlier point about the
reputation threat from data breaches.

Besides the odd paper or two, there appears to be no explicitly related previous
works. Implicitly, however, the paper has connections to many directions, as
indicated by the points raised and the works referenced in the previous
section. Of these directions, the so-called event studies are noteworthy because
these have used data breach news as an explanatory factor in explaining firms'
economic performance~\cite{Spanos16}. If the scope would be enlarged to crime
news and crime journalism more generally, also the amount of related works would
enlarge. Regarding cyber crime and particularly data breaches, however, the
amount of related works is minimal.


\section{Data and Methods}\label{sec: data and methods}

\subsection{Data}\label{subsec: data}

The dataset was assembled by web scraping \textit{databreaches.net}'s
content. The category of breach incidents was used for downloading the news
articles. There were $19,348$ news articles in this category during the data
retrieval in April 21, 2024. However, not all of these articles are about data
breaches; there are plenty of non-breach cyber crime and general cyber security
news in the category as well. Therefore, as a case-specific solution, only those
news articles were further processed whose lower-cased titles contained the
character string \texttt{breach}, \texttt{leak}, \texttt{steal}, or
\texttt{stole}. The latter two strings are used to cover the various cases
whereby either employees have stolen data or computers and the data stored in
them were stolen.

Due to the various ways data breaches can occur, the four strings do not capture
all cases, of course. For instance, \textit{databreaches.net} has reported also
on the several cases of data being found on dumpsters or other places, the cases
where employees have accidentally sent personal data by email or posted the data
on social media, and so on. However, according to some studies, only a minority
of breaches is due to inside jobs; the majority is due to external, remote
technical exploitation, and lost or stolen equipment together attain the second
place~~\cite{Phua09}. Although also unauthorized disclosures have been common in
some sectors~\cite{Lee23}, the two typical cases are well-covered; technical
exploitation and more traditional theft. A~portion of unauthorized disclosures
is additionally captured by the character string \texttt{leak}. The resulting
sample size of $n = 5,861$ news articles is also more than sufficient for the
paper's exploratory purposes.

\subsection{Methods}

The overall methodology taken is an exploratory case study~\citep{Gerring04}. In
other words, the goal is to explore the dataset without strict prior hypothesis,
providing a rich detailed study on the particular case, generating hypothetical
paths for further work, and in this way advancing the previously unexplored
field of study. As the name indicates, \textit{databreaches.net} concentrates on
data breaches, and, hence, it provides also an ideal case to study data breach
journalism. That said, the methodology has its own limitations, which are
discussed later on in Section~\ref{subsec: limitations}. For the present
purposes, it is more important to focus on a few practical computational
details.

Most of the quantified information is extracted by conventional parsing of the
meta-data or the content of the news articles. Instead of elaborating this
everyday computing, a few words must be written about the ways the content of
the articles is summarized. Instead of relying on topic modeling, the most
frequent ``topics'' are briefly probed and summarized with bigrams (that is, two
consecutive character strings). To construct the bigrams, the following steps
were used after having lower-cased and tokenized the body of the news articles
according to white space and punctuation characters: only tokens representing
alphabetical character strings recognized as English words were included, tokens
with length less than three or larger than twenty characters were excluded, and
tokens matching stopwords were also excluded. After this pre-processing, the
tokens were lemmatized into their basic English dictionary forms. Two libraries
were used for this processing~\cite{NLTK24, hunspell24}.\footnote{~In addition
  to the stock stopwords in the library used \cite{NLTK24}, the following
  stopwords were used: \textit{data}, \textit{breach}, \textit{story},
  \textit{new}, \textit{newly}, \textit{update}, \textit{case}, \textit{posted},
  \textit{reported}, \textit{information}, \textit{web}, \textit{site},
  \textit{may}, and \textit{must}.} In addition to this simple bigram
processing, sentiment analysis is used to probe for the potential sensationalism
in the news articles.

The lexicon and rule based VADER algorithm is used for computing the
sentiments~\cite{VADER}. There are a few things bespeaking for the algorithm in
the present context. Namely, it is simple, easily available, and hugely
popular. Despite the algorithm's simplicity, it also performs well in
benchmarks~\cite{Ribeiro16}. Regarding popularity, on the other hand, there are
literally thousands of studies (if not more) that have used the algorithm in
various contexts. For the present purposes, it suffices to note that the
algorithm has been used to examine the sentiments in news
articles~\cite{Castellanos21}. Whatever the application context, the algorithm
assigns positive and negative values to words, and then combines the values that
appear both in the VADER's lexicon and the given input text. The final score is
normalized into a range $[-1, 1]$. In light with the discussion in
Section~\ref{subsec: related work}, negative values, meaning negative
sentiments, could be expected for much of the news articles. After all, crime is
a negative phenomenon and sensationalist crime news have a long history.

\section{Results}\label{sec: results}

\subsection{Sample Characteristics}

\begin{figure}[th!b]
\centering
\includegraphics[width=\linewidth, height=2.9cm]{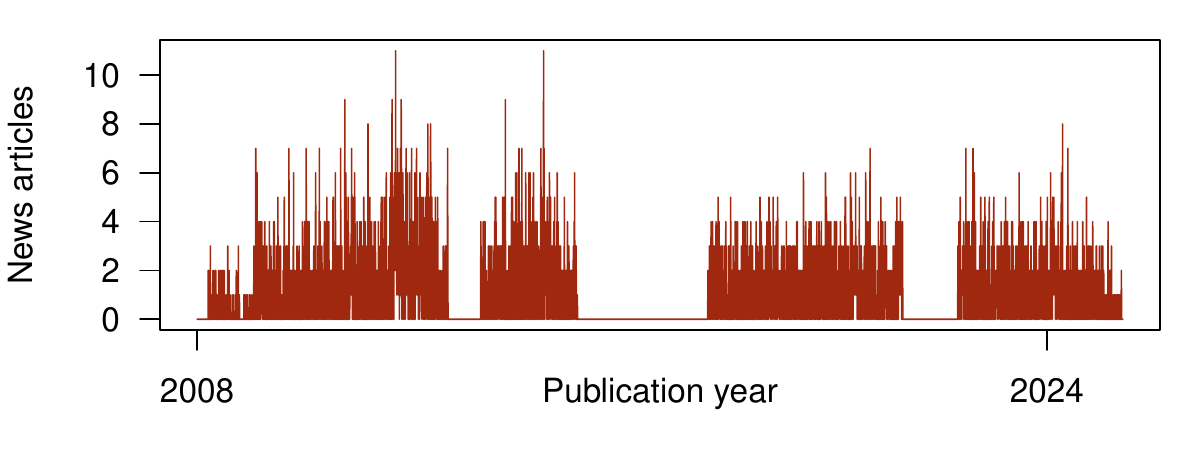}
\caption{Publication Dates}
\label{fig: dates}
\end{figure}

Four points about the sample are a good way to start the empirical
exploration. Thus, first, Fig.~\ref{fig: dates} shows the number of news
articles published across the time period observed. As can be seen, the outlet
has kept a steady publishing cycle; on average, about two news articles were
published each day. That said, there have been some notable gaps in the
publishing over the years. As a small outlet, a lack of sufficient human
resources may explain these gaps.

Regarding time series trends in general, it has been hypothesized and observed
that recent events and transformations, such as \text{COVID-19} and increased
remote work, may have increased the amount of data
breaches~\cite{Ozer24}. However, such conclusions are uncalled for with the
present case study; data breaches are one thing and journalistic reporting on
them another. There is also another point worth making about the longitudinal
aspects: data breaches take a long time to be detected. According to some
studies and industry reports, even more than a hundred days is typically
required before a data breach is detected within an organization~\cite{IBM23,
  Roumani22}. If no leaks occur to journalists, supposedly a long time is
further required before news about a breach appear in media. Thus, the
publication dates shown do not likely say much about the actual dates of the
data breaches discussed in the news articles.

\begin{figure}[th!b]
\centering
\includegraphics[width=\linewidth, height=3.5cm]{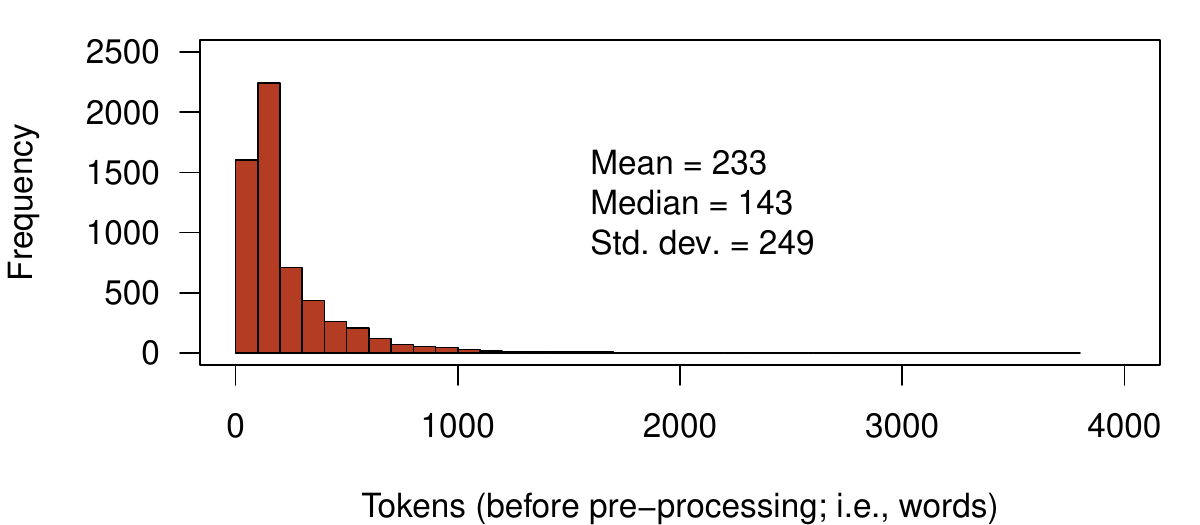}
\caption{Length of the News Articles}
\label{fig: unigrams}
\end{figure}

Second, as can be concluded from Fig.~\ref{fig: unigrams}, the news articles are
typically very short. In fact, the mean length, as measured by the number of
tokens (that is, words), is about the same as the length of a typical abstract
in a scientific paper. Although also longer news articles are present, it can be
concluded that the typical articles do not perhaps have much journalistic
ambition. Instead, these focus on plain reporting on data breaches, short and
simple.

\begin{figure}[th!b]
\centering
\includegraphics[width=\linewidth, height=8cm]{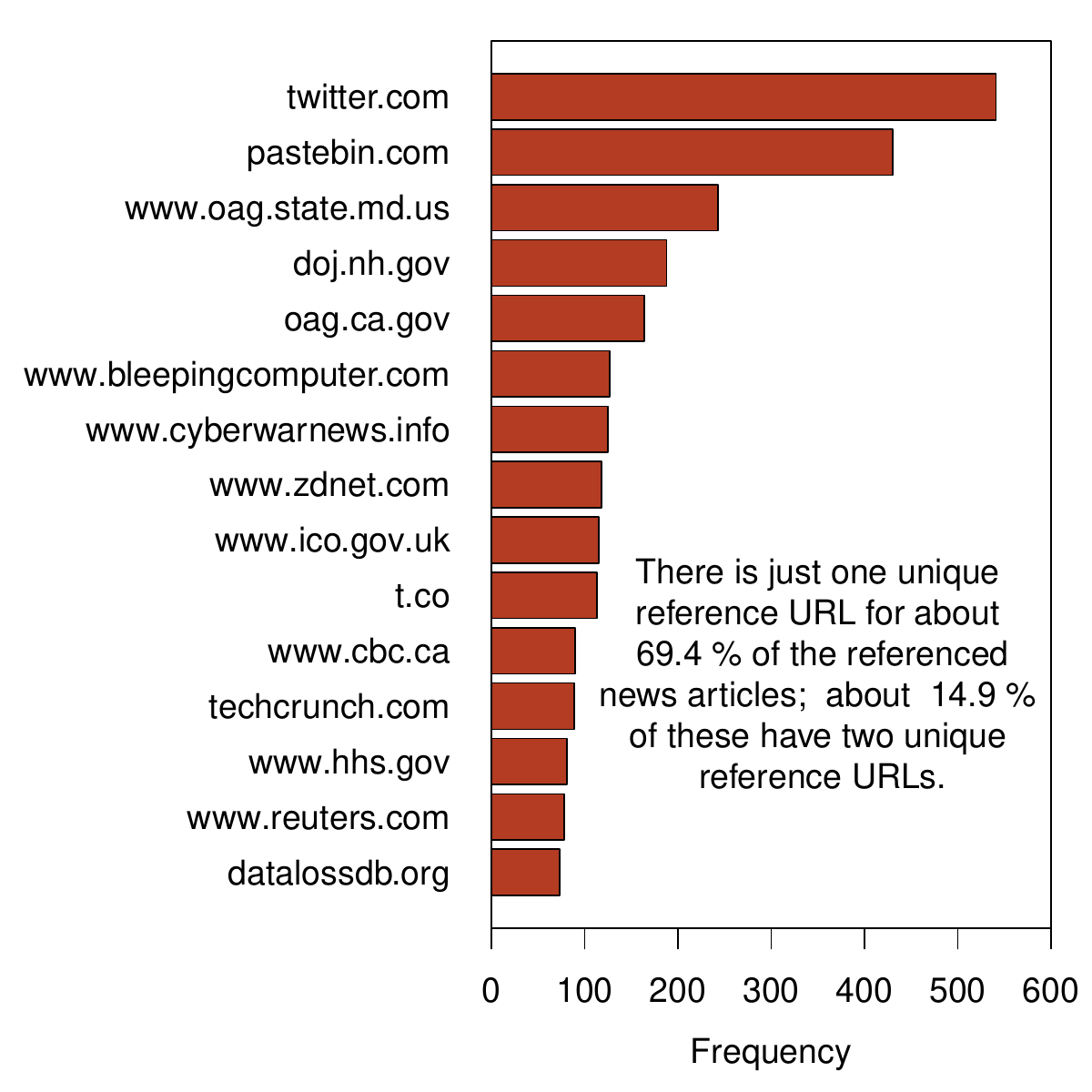}
\caption{Top-15 Domains in the Articles' URL References}
\label{fig: domains}
\end{figure}

\begin{figure}[th!b]
\centering
\includegraphics[width=\linewidth, height=2.9cm]{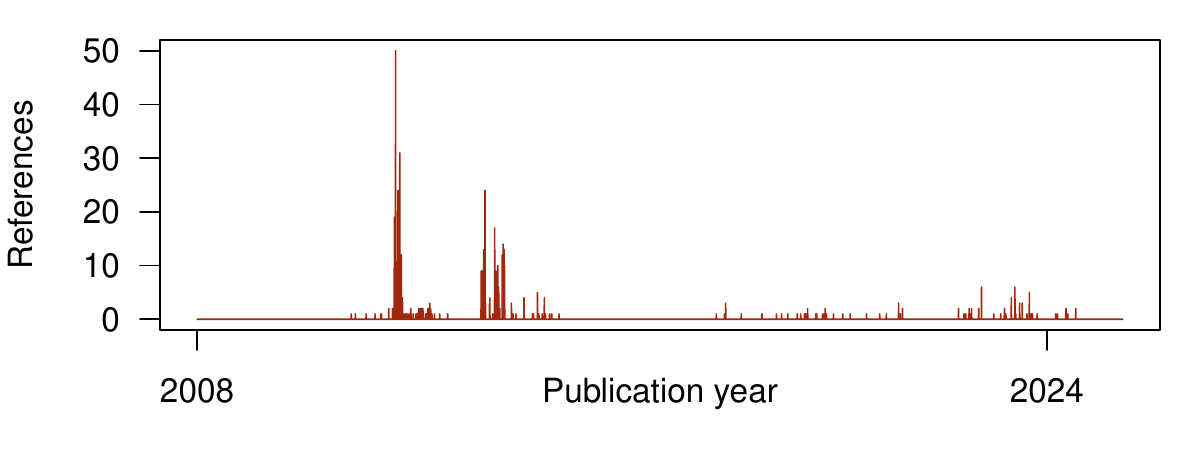}
\caption{Publication Dates of Twitter References}
\label{fig: twitter}
\end{figure}

Third, the site offers commenting functionality but it is not much used. In
fact, only about 5\% of the news articles in the sample have seen at least one
comment. While this observation cannot be equated to popularity, it can be still
concluded that the site itself does not rely on social aspects for its reporting.

Fourth, a few interesting observations can made based on the sources on which
the news articles are based. A simple way to approximate the sources is to focus
on the domain names that appear in the uniform resource locators (URLs) embedded
to hyperlinks in the body of the news articles, excluding self-references to
other news articles on \textit{databreaches.net}. The top-15 most referenced
domains are shown in Fig.~\ref{fig: domains}. The most notable observation is
present in the text embedded to the middle of the figure; most cases are
referenced with a single unique URL. This observation hints that most of the
cases are referenced with primary sources that likely point toward the victim
organizations. Although such referencing clearly increases quality, there are
also quite a few references to X/Twitter. It may not be a big sin, however,
given that also mainstream media nowadays tend to often reference this social
media platform. It may also be that X/Twitter has gained prominence in data
breach reporting. However, according to Fig.~\ref{fig: twitter}, most of the
references were made in the late 2000s during which also X/Twitter's popularity
peaked. Otherwise, a commonplace English language bias seems to be present,
which is expected. Technology news sites in English, the public broadcaster from
Canada, the data protection authority in the United Kingdom, and several
governmental websites in the United States have been relatively frequently
referenced in the news articles. All are quality sources. Finally, it should be
also remarked that the site occasionally uses also traditional journalistic
referencing, mentioning a source for a news article in plain text, without
providing associated hyperlinks to the online origins.

\subsection{Topics}

\begin{figure}[th!b]
\centering
\includegraphics[width=\linewidth, height=11.0cm]{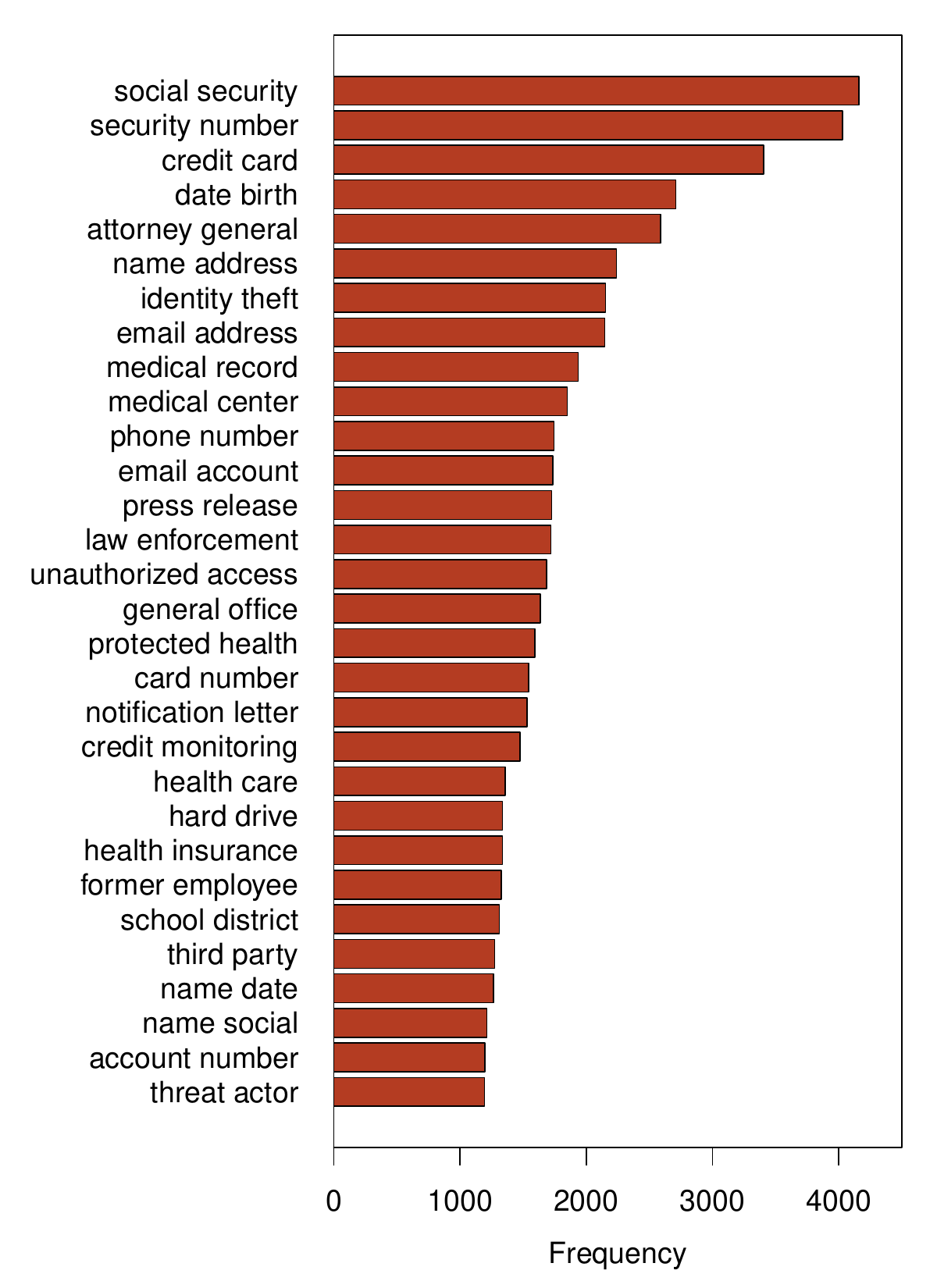}
\caption{Top-30 Bigrams}
\label{fig: bigrams}
\end{figure}

The bigrams provide sufficient material to deduce abut the typical cases
discussed in the data breach news articles. To this end, Fig.~\ref{fig: bigrams}
displays the top-25 bigrams present in the pre-processed sample. In addition,
Table~\ref{tab: sectors} shows the sectors of the victim organizations according
to existing meta-data provided by the news site. None of the sectors exhibit
notable longitudinal trends; the data breaches have affected the five sectors
rather uniformly across time.

\begin{table}[th!b]
\centering
\caption{Sectors According to Meta-Data}
\label{tab: sectors}
\begin{tabular}{lc}
\toprule
Sector & \% of breaches \\
\hline
Healthcare & 33.3 \\
Business & 26.7 \\
Government & 16.6 \\
Education & 10.9 \\
Financial & 5.4 \\
\hline
Other & 7.1 \\
\bottomrule
\end{tabular}
\end{table}

The following points can be used to summarize the results:
\begin{itemize}

\itemsep 3pt

\item{Many of the breaches have involved data that can be used to either
  directly or non-directly identify natural persons. The examples include social
  security numbers, dates of birth, email and street addresses, and phone
  numbers.}

\item{Commercial data is well-represented; according to the meta-data, the financial
  and general business sectors together account for about 32\% of the breaches
  reported. According to the bigrams, the breaches have often involved credit
  card details. Also credit monitoring organizations are covered. Furthermore,
  supply chain attacks or crimes are present, given that 8\% of the cases have
  involved subcontractors according to the existing meta-data provided by the site.}

\item{Due to the nature of the data breached, and despite the sampling strategy
  (see Section~\ref{subsec: data}), the news articles have also addressed
  related crime, as manifested by the bigrams \texttt{identity theft},
  \texttt{attorney general}, and \texttt{law enforcement}. In addition, the
  bigrams \texttt{former employee} and \texttt{hard drive} hint that not all of
  the data breaches have occurred because of technical exploitation, as was
  expected. Indeed, according to the meta-data, about 9\% of the data breaches
  have been inside jobs and about 8\% have involved stolen equipment.}

\item{The healthcare sector is represented particularly well. There are also
  many bigrams explicitly related to the sector, including, but not limited to,
  \texttt{medical record}, \texttt{medical center}, \texttt{protected health},
  and \texttt{health insurance}.}

\item{In addition to healthcare, which may have public sector ties, the public
  sector is also otherwise well-represented, as hinted by the bigram
  \texttt{school district}. In fact, together the education and government
  sectors account for about 28\% of the data breaches according to the available
  meta-data.}

\item{Although information security incidents have been a common cause for GDPR
  fines~\cite{Ruohonen22IS}, only seventeen news articles have contained the
  character string \texttt{gdpr}. This observation likely reflects the English
  language bias and the origins of the news site in the United States. Indeed,
  the character string \texttt{hipaa}, as for the Health Insurance Portability
  and Accountability Act (HIPAA) in the United States, already accounts for 221
  mentions in the data breach news articles. Also the California Consumer
  Privacy Act (CCPA), which was motivated by the GDPR, has gathered one
  reference.}

\end{itemize}

All in all, the most frequent bigrams indicate that many of the cases are bread
and butter data breaches. However, this mundane nature of many of the breaches
should not be taken to undermine the risks to the victims. As many---if not the
most---breaches have involved direct personal identifiers, identity theft is a
real risk for the victims. This risk is further increased by the widespread
leakage of credit card and credit ranking details. Alongside with the discussion
in Section~\ref{subsec: consequences}, it can be also remarked that according to
some studies, some consumers rely on fraud prevention services after having
become aware of a breach but these people still use existing credit cards, and
data breach news have generally only a limited impact upon consumer
behavior~\cite{Mikhed18}. Thus, the risks are real in the sense that these may
involve financial losses, lowering of credit ratings, or something
similar. Regarding financial or other crime, it can be also remarked that about
16\% of the news articles have contained a character string from the list
\texttt{detain}, \texttt{criminal}, \texttt{suspect}, \texttt{jail},
\texttt{prison}, and \texttt{fine}. Although this detail again pinpoints that
the sampling strategy is not perfect, it also serves to underline that many of
the cases are about serious crimes and security lapses. The appearance of these
strings may be amplified by the frequent sourcing of the material from law
enforcement, public attorneys, data protection authorities, and related official
figures.

The healthcare data breaches warrant a brief discussion too. Although financial
crime may be involved due to stolen health insurance details, the risks
associated are likely different for many victims. As was already noted, leakage
of patient records may cause psychological stress to many victims. These can be
used also on more serious crimes, such as blackmailing schemes. In terms of the
victim organizations, on the other hand, healthcare data breaches are
particularly costly for the organizations involved~\cite{IBM23}. Another big
question is why hospitals and healthcare organizations in general are so prone
to data breaches and cyber attacks in general. Alas, the reasons are
many. Security awareness, human competencies, organizational postures, and
financial resources are problems, but these are further exacerbated by lack of
oversight bodies and poor governance of healthcare systems~\cite{GarziaPerez23,
  Lee23}, among other things. For the time being, it seems sensible to expect
widespread insecurity of hospitals and healthcare facilities also in the
foreseeable future. The data breaches presumably do not stop in this sector, no
matter how difficult the consequences are for the victims of the~breaches.

\subsection{Sentiments}

\begin{figure}[th!b]
\centering
\includegraphics[width=\linewidth, height=8cm]{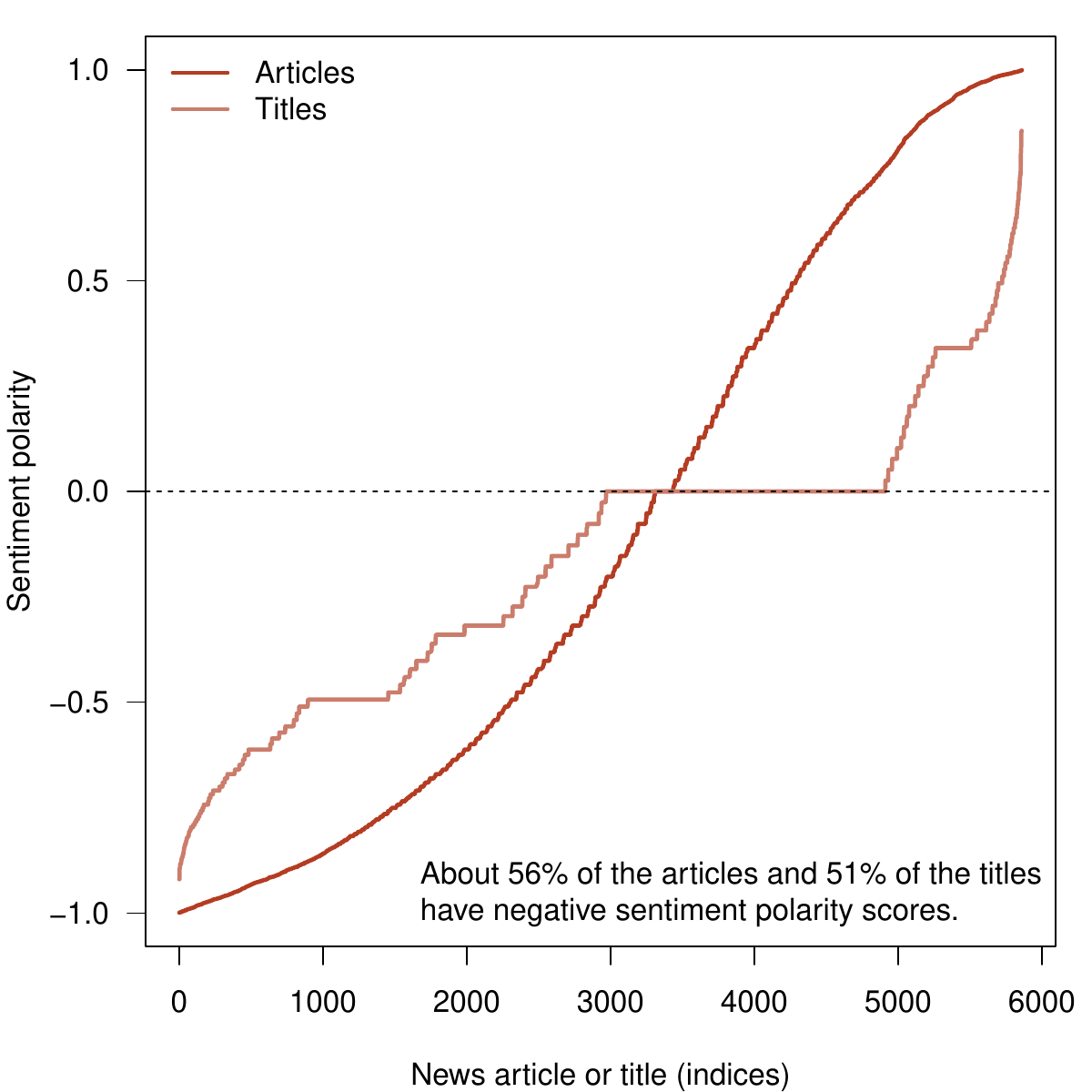}
\caption{Sentiments}
\label{fig: sentiments}
\end{figure}

Sentiment analysis is the final step in the empirical exploration. Thus,
Fig.~\ref{fig: sentiments} displays the sentiment polarity scores for the news
articles and their titles. Unlike what was expected and contrary to existing
research results~\cite{Sinanaj14}, the news articles split almost perfectly into
half between negative and positive polarity scores. Moreover, there are not many
articles with values around zero, indicating that the news articles tend to
exhibit relatively strong negative or positive sentiments. This observation is
even more surprising when consider the short length of the articles. Emotionally
loaded language provides one explanation, although it is difficult to speculate
why positive sentiments have been common in the articles. In contrast, the
articles with negative sentiment polarity scores are presumably at least
partially due to the relatively widespread appearance of crime-related topics in
the news articles. Moreover, as can be seen from Fig.~\ref{fig: sentiments
  time}, the sentiment polarity scores have remained stable across time. Thus,
the negative or positive sentiments cannot be attributed to a sudden change in
the news site's reporting style.

\begin{figure}[th!b]
\centering
\includegraphics[width=\linewidth, height=2.9cm]{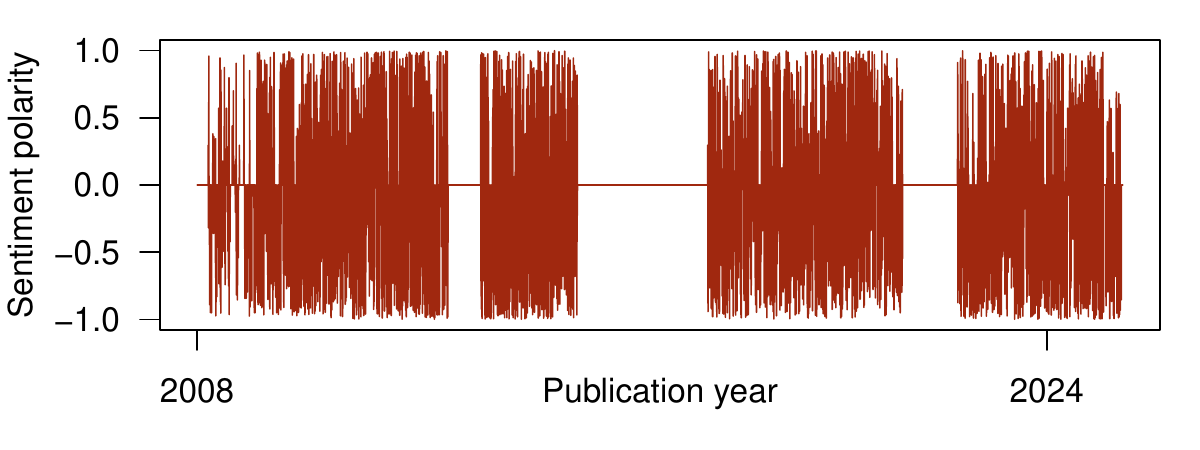}
\caption{Sentiments Across Publication Dates}
\label{fig: sentiments time}
\end{figure}

The titles of the articles exhibit a relatively large volume at the value zero,
meaning no sentiment tendency to either direction. Even then, however, also the
titles tend to split between negative and positive sentiments. These may
indicate the presence of so-called clickbaits, although a closer examination can
be left to further work.

\section{Discussion}\label{sec: discussion}

\subsection{Conclusion}

This paper investigated the mostly unexplored topic of data breach journalism
with an exploratory case study of \textit{databreaches.net}. The paper's results
can be summarized with the following points:
\begin{itemize}

\itemsep 3pt

\item{The news site has kept a steady publishing pace, although there have also
  been some gaps in the publishing over the years. On average, two data breach
  news articles have appeared on each day after the site's launch in 2008.}

\item{The news articles are very short; the site concentrates on plain reporting
  instead of feature journalism or other more in-depth investigations.}

\item{The news are often sourced with primary material from the victim
  organizations.  Although also social media is present, additional news
  material includes high-quality sources, including technology news sites,
  criminal justice systems, and data protection authorities. That said,
  practically all of the sources are in English, which likely reflects a
  geographic~bias.}

\item{The full scale of data breaches has been covered. However, many of the
  breaches discussed by the news site are fairly traditional, exposing direct
  personal identifiers and financial details of the victims. In addition,
  hospitals and the healthcare sector stand out as frequent victims of data
  breaches. The breaches are often discussed in relation to cyber crime.}

\item{The news articles exhibit fairly strong positive and negative
  sentiments. The negative ones are expected and understandable due to the
  presence of crime and the long history of sensationalism in crime news, but
  the positive sentiments are unexpected and would require further
  evaluation. The titles of the news articles are more balanced, although also
  these show a split into negative and positive sentiments, possibly indicating
  the presence of so-called clickbaits.}

\end{itemize}

In addition to this short summary of the results, a few limitations should be
acknowledged. After outlining these limitations, topics and paths for further
research are briefly discussed.

\subsection{Limitations}\label{subsec: limitations}

A notable limitation stems from the case study approach: there are no other
cases to which the results could be compared~\citep{Gerring04}. Specifically,
many of the results would benefit from a comparison to handling of data breaches
in mainstream media. Such a comparison would be particularly interesting with
respect to the sentiment analysis and the crime characteristics; by hypothesis,
there is a difference, although it is difficult to say to which
direction. Coverage of mainstream media would be important also in terms of the
discussion in~Sections~\ref{subsec: crime news} and \ref{subsec:
  consequences}. In other words, \textit{databreaches.net}'s readership is
presumably rather narrow, supposedly mainly composed of information technology
professionals, whereas the readership of mainstream media outlets is wide; by
implication, many of the issues in crime journalism supposedly only manifest
themselves through mainstream media. The same applies to the consequences
hypothesized about data breach journalism. Also a few smaller limitations should
be briefly mentioned.

One smaller limitation originates from the sampling strategy (see
Section~\ref{subsec: data}). Notably, the inclusion of news articles with the
string \texttt{breach} in the articles' titles covers also cases that have
discussed politics, law, and regulations, which are somewhat oddly often
entitled as ``breach regulations''. However, it is not easy to address this
limitation: if the strings \texttt{regulation} and \texttt{law}, in turn, would
be used to exclude news articles, many relevant news articles would be missed,
as the news site has often discussed breaches in conjunction with regulations. A
similar point applies to crime news, which, as was noted, still appear in the
sample despite the sampling strategy. Therefore, a bias must be acknowledged:
not all of the news articles covered are only about data breaches \textit{per~se}.

A small limitation is present also in terms of follow-up news that update a
given article. Although it is difficult to say how many such follow-up stories
are present in the sample, these duplicates may cause a small bias in the
results reported. A somewhat different bias stems from potential incorrect
reporting to which a niche news site may be subject to. Unlike major mainstream
media outlets, \textit{databreaches.net} appears to avoid publishing corrections
to previous articles. Nor does it publish potential retractions in case of
severely misleading reporting. However, the resulting biases are likely small
and unavoidable in the present case study context; deducing about these would
require the examination of other media sources.

Finally, limitations of the sentiment analysis algorithm should be briefly
acknowledged. The VADER algorithm was developed with social media analytics in
mind. Hence, its lexicon may not be the optimal choice in the present context,
which might require a custom dictionary tailored specifically for data breach
news. This point generalizes toward other, newer sentiment analysis
algorithms. Also the other common limitations apply, including sentiment
analysis algorithms' poor handling of adverbs, irony, sarcasm, context, and
comparative sentences~\cite{Bordoloi23, Denecke23}. Sentiment analysis
algorithms are also sensitive to different pre-processing choices. In the
present context these choices include the sampling strategy in particular.

\balance
\subsection{Further Work}

A few promising paths for further research can be mentioned. As was already noted
in the previous section, the scope of the analysis should be extended toward
mainstream media. Such an extension would allow to better understand the results
as there would be a comparative case against which the results could be
reflected. In particular, the sentiment analysis results would benefit from a
comparison; it may be that mainstream media is more prone to sensationalism due
to its coverage of traditional crime news as well. Though, there are likely
differences among mainstream media outlets as well; the sentiments may be quite
different among something like \textit{The Guardian} and \textit{The
  Register}. A comparison could be done also between the sensationalism (or lack
thereof) of cyber crime news and traditional crime news. Furthermore, it would
be interesting to know how much data breaches covered by the dedicated outlet
are missed in mainstream media due to length and coverage constraints. As the
outlet studied does not use images, meainstream media coverage could be used to
examine also the visual cues, which are a distinct trait of crime
journalism~\cite{Higgins22}. In addition, coverage in mainstream media is better
suited to examine correlations to people's security awareness, among other
things.

Another interesting path for further work would involve separating the distinct
victim organizations from the data breach news. Although such separation may
require manual inspection of the news articles, the resulting data would offer
material for many new research questions. For instance, it would be possible to
continue the event study approaches by incorporating the results from sentiment
analysis as explanatory factors. In a similar vein, it would be interesting to
combine resolved data breach crimes, from the initial reporting of the breaches
to the actions taken by a criminal justice system. Such a combination would
allow to better understand the criminological aspects of cyber crime related to
data breaches.

Sentiment analysis could be also used in a slightly different context by
modeling the public's responses to data breaches. According to existing results
based on sentiment analysis of social media data, the responses tend to exhibit
distinct periods marked by anxiety, anger, sadness, and
acceptance~\cite{Bachura17}. Also fear has been observed to be a typical
emotional response~\cite{Chatterjee19}. Media is likely facilitating these
distinct periods; hence, it would be interesting to see whether the sentiments
in news articles and social media data correlate.

Finally, it is possible that sentiment analysis is a poor tool to examine
longer, more nuanced, and contextualized texts. To this end, qualitative
analysis could be carried out to better understand the style, linguistic
conventions, and semantics of data breach news and cyber crime news in
general. As these supposedly constitute a distinct news genre of their own, it
may be that also the linguistic conventions and stylistic elements are distinct
among news genres.



\clearpage
\newpage
\balance
\bibliographystyle{abbrv}

\begin{thebibliography}{10}

\bibitem{Bachura17}
E.~Bachura, R.~Chen, R.~Valecha, and H.~R. Rao.
\newblock {M}odeling {P}ublic {R}esponse to {D}ata {B}reaches.
\newblock In {\em Proceedings of the Twenty-Third Americas Conference on
  Information Systems}, Boston, 2017. AIS.
\newblock {A}vailable online in April, 2024:
  \url{https://core.ac.uk/download/pdf/301372705.pdf}.

\bibitem{Bhagavatula21}
S.~Bhagavatula, L.~Bauer, and A.~Kapadia.
\newblock {W}hat {B}reach? {M}easuring {O}nline {A}wareness of {S}ecurity
  {I}ncidents by {S}tudying {R}eal-{W}orld {B}rowsing {B}ehavior.
\newblock In {\em Proceedings of the 2021 European Symposium on Usable Security
  (EuroUSEC 2021)}, pages 180--199, Karlsruhe, 2021. ACM.

\bibitem{Boholm21}
M.~Boholm.
\newblock {T}wenty-{F}ive {Y}ears of {C}yber {T}hreats in the {N}ews: {A}
  {S}tudy of {S}wedish {N}ewspaper {C}overage (1995--2019).
\newblock {\em Journal of Cybersecurity}, 7(1):1--23, 2021.

\bibitem{Bordoloi23}
M.~Bordoloi and S.~K. Biswas.
\newblock {S}entiment {A}nalysis: {A} {S}urvey on {D}esign {F}ramework,
  {A}pplications and {F}uture {S}copes.
\newblock {\em Artificial Intelligence Review}, 56:12505--12560, 2023.

\bibitem{Castellanos21}
E.~Castellanos, H.~Xie, and P.~Brenner.
\newblock {G}lobal {N}ews {S}entiment {A}nalysis.
\newblock In Z.~Yang and E.~{von Briesen}, editors, {\em {P}roceedings the 2019
  {I}nternational {C}onference of the {C}omputational {S}ocial {S}cience
  {S}ociety of the {A}mericas}. Springer, 2021.
\newblock {S}pringer {P}roceedings in {C}omplexity.

\bibitem{Chatterjee19}
S.~Chatterjee, X.~Gao, S.~Sarkar, and C.~Uzmanoglu.
\newblock {R}eacting to the {S}cope of a {D}ata {B}reach: {T}he {D}ifferential
  {R}ole of {F}ear and {A}nger.
\newblock {\em Journal of Business Research}, 101:183--193, 2019.

\bibitem{Christian22}
J.~Christian, S.~Valiveti, and S.~Jain.
\newblock {P}rofiling {C}yber {C}rimes from {N}ews {P}ortals {U}sing {W}eb
  {S}craping.
\newblock In P.~K. Singh, S.~T. Wierzcho\'n, J.~K. Chhabra, and S.~Tanwar,
  editors, {\em {F}uturistic {T}rends in {N}etworks and {C}omputing
  {T}echnologies}, pages 1007--1016. Springer, Singapore, 2022.
\newblock {S}elect {P}roceedings of {F}ourth {I}nternational {C}onference on
  {FTNCT} 2021, {L}ecture {N}otes in {E}lectrical {E}ngineering 936.

\bibitem{Chua21}
H.~N. Chua, J.~S. Teh, and A.~Herbland.
\newblock {I}dentifying the {E}ffect of {D}ata {B}reach {P}ublicity on
  {I}nformation {S}ecurity {A}wareness {U}sing {H}ierarchical {R}egression.
\newblock {\em IEEE Access}, 9:121759--12770, 2021.

\bibitem{Denecke23}
K.~Denecke.
\newblock {\em {S}entiment {A}nalysis in the {M}edical {D}omain}.
\newblock Springer, Cham, 2023.

\bibitem{Doyle06}
A.~Doyle.
\newblock {H}ow {N}ot to {T}hink about {C}rime in the {M}edia.
\newblock {\em Canadian Journal of Criminology and Criminal Justice},
  48(6):867--885, 2006.

\bibitem{GarziaPerez23}
A.~{Garcia-Perez}, J.~G. {Cegarra-Navarro}, M.~P. Sallos, E.~{Martinez-Caro},
  and A.~Chinnaswamy.
\newblock {R}esilience in {H}ealthcare {S}ystems: {C}yber {S}ecurity and
  {D}igital {Y}ransformation.
\newblock {\em Technovation}, 121:102583, 2023.

\bibitem{Gerring04}
J.~Gerring.
\newblock {W}hat {I}s a {C}ase {S}tudy and {W}hat {I}s {I}t {G}ood for?
\newblock {\em American Political Science Review}, 98(2):341--354, 2004.

\bibitem{Gibson23}
D.~Gibson and C.~Harfield.
\newblock {A}mplifying {V}ictim {V}ulnerability: {U}nanticipated {H}arm and
  {C}onsequence in {D}ata {B}reach {N}otification {P}olicy.
\newblock {\em International Review of Victimology}, 29(3):341--365, 2023.

\bibitem{Higgins22}
K.~C. Higgins.
\newblock {R}ethinking {V}isual {C}riminalization: {N}ews {I}mages and the
  {M}ediated {S}pacetime of {C}rime {E}vents.
\newblock {\em Visual Communication}, (Published online in June):1--23, 2022.

\bibitem{Huey12}
L.~Huey, J.~Nhan, and B.~Broll.
\newblock ‘{U}ppity {C}ivilians’ and ‘{C}yber-{V}igilantes’: {T}he
  {R}ole of the {G}eneral {P}ublic in {P}olicing {C}yber-{C}rime.
\newblock {\em Criminology \& Criminal Justice}, 13(1):81--97, 2012.

\bibitem{VADER}
C.~J. Hutto and E.~Gilbert.
\newblock {VADER}: {A} {P}arsimonious {R}ule-{B}ased {M}odel for {S}entiment
  {A}nalysis of {S}ocial {M}edia {T}ext.
\newblock In {\em Proceedings of the Eighth International AAAI Conference on
  Weblogs and Social Media}, pages 216--225, Ann Arbor, 2014. AAAI.

\bibitem{IBM23}
{IBM}.
\newblock {C}ost of a {D}ata {B}reach {R}eport.
\newblock {A}vailable online in June 2024:
  \url{https://www.ibm.com/reports/data-breach}, 2023.

\bibitem{Katz87}
J.~Katz.
\newblock {W}hat {M}akes {C}rime '{N}ews'?
\newblock In C.~Greer, editor, {\em Crime and Media}, pages 47--75. Routledge,
  London, 1987.

\bibitem{Lavorgna19}
A.~Lavorgna.
\newblock {C}yber-{O}rganised {C}rime. {A} {C}ase of {M}oral {P}anic?
\newblock {\em Trends in Organized Crime}, 22:357--374, 2019.

\bibitem{Lee23}
I.~Lee.
\newblock {A}nalyzing {W}eb {D}escriptions of {C}ybersecurity {B}reaches in the
  {H}ealthcare {P}rovider {S}ector: {A} {C}ontent {A}nalytics {R}esearch
  {M}ethod.
\newblock {\em Computers \& Security}, 129:103185, 2023.

\bibitem{Lindgren08}
S.~Lindgren.
\newblock {C}rime, {M}edia, {C}oding: {D}eveloping a {M}ethodological
  {F}ramework for {C}omputer-{A}ided {A}nalysis.
\newblock {\em Crime, Media, Culture: An International Journal}, 4(1):95--100,
  2008.

\bibitem{Madnick24}
S.~Madnick.
\newblock {W}hy {D}ata {B}reaches {S}piked in 2023, 2024.
\newblock {H}arvard {B}usiness {R}eview. {A}available online in April:
  \url{https://hbr.org/2024/02/why-data-breaches-spiked-in-2023}.

\bibitem{Mayer23}
P.~Mayer, Y.~Zou, B.~M. Lowens, H.~A. Dyer, K.~Le, F.~Schaub, and A.~J. Aviv.
\newblock {A}wareness, {I}ntention, {(In)}{A}ction: {I}ndividuals' {R}eactions
  to {D}ata {B}reaches.
\newblock {\em ACM Transactions on Computer-Human Interaction}, 30(5):77:1 --
  77:53, 2023.

\bibitem{Mazrouei22}
M.~K.~A. Mazrouei, D.~C. Piper, and L.~Y. Connolly.
\newblock {B}eware of {T}itles: {A}nalysing {M}edia {R}eporting of {C}ybercrime
  in {UK} and {UAE}.
\newblock In {\em Proceedings of the Cyber Research Conference -- Ireland
  (Cyber-RCI 2022)}, pages 1--5, Galway, 2022. IEEE.

\bibitem{Mikhed18}
V.~Mikhed and M.~Vogan.
\newblock {H}ow {D}ata {B}reaches {A}ffect {C}onsumer {C}redit.
\newblock {\em Journal of Banking and Finance}, 88:192--207, 2018.

\bibitem{hunspell24}
L.~N\'emeth, K.~Hendricks, C.~McNamara, et~al.
\newblock {Hunspell}.
\newblock {V}ersion 1.7.0, available online in February
  \url{https://github.com/hunspell/hunspell}, 2024.

\bibitem{OLeary21}
N.~O'Leary.
\newblock {\em {A} {V}ictim {C}ommunity: {S}tigma and the {M}edia {L}egacy of
  {H}igh-{P}rofile {C}rime}.
\newblock Palgrave Macmillan, Cham, 2021.

\bibitem{Ozer24}
M.~Ozer, Y.~Kose, M.~Bastug, G.~Kucukkaya, and E.~R. Varlioglu.
\newblock {T}he {S}hifting {L}andscape of {C}ybersecurity: {T}he {I}mpact of
  {R}emote {W}ork and {COVID-19} on {D}ata {B}reach {T}rends.
\newblock {A}rchived manuscript. Available online in April:
  \url{https://arxiv.org/abs/2402.06650}, 2024.

\bibitem{Parks19}
P.~Parks.
\newblock {N}aturalizing {N}egativity: {H}ow {J}ournalism {T}extbooks {J}ustify
  {C}rime, {C}onflict, and ``{B}ad'' {N}ews.
\newblock {\em Critical Studies in Media Communication}, 36(1):75--91, 2019.

\bibitem{Phua09}
C.~Phua.
\newblock {P}rotecting {O}rganisations from {P}ersonal {D}ata {B}reaches.
\newblock {\em Computer Fraud \& Security}, 1:13--18, 2009.

\bibitem{Reitberger17}
G.~Reitberger and S.~Wetzel.
\newblock {I}nvestigating the {I}mpact of {M}edia {C}overage on {D}ata {B}reach
  {F}atigue.
\newblock In {\em Proceedings of the IEEE 38th Sarnoff Symposium}, pages 1--3,
  Newark. IEEE.

\bibitem{Ribeiro16}
F.~N. Ribeiro, M.~Ara\'ujo, P.~Gon\c{c}alves, M.~A. Gon\c{c}alves, and
  F.~Benevenuto.
\newblock {S}enti{B}ench -- {A} {B}enchmark {C}omparison of
  {S}tate-of-the-{P}ractice {S}entiment {A}nalysis {M}ethods.
\newblock {\em EPJ Data Science}, 5:1--29, 2016.

\bibitem{Rice29}
T.~R. Rice.
\newblock {H}ow to {M}ake {C}rime {N}ews {C}onstructive.
\newblock {\em The Journalism Quarterly}, 5(4):18--22, 1929.

\bibitem{Rios20}
V.~Rios and C.~J. Ferguson.
\newblock {Ne}ws {M}edia {C}overage of {C}rime and {V}iolent {D}rug {C}rime:
  {A} {C}ase for {C}ause or {C}atalyst?
\newblock {\em Justice Quarterly}, 37(6):1012--1039, 2020.

\bibitem{Robertson23}
C.~E. Robertson, N.~Pr\"ollochs, K.~Schwarzenegger, P.~P\"arnamets, J.~J. {van
  Bavel}, and S.~Feuerriegel.
\newblock {N}egativity {D}rives {O}nline {N}ews {C}onsumption.
\newblock {\em Nature Human Behavior}, 7:812--822.

\bibitem{Rossland07}
L.~A. R\o{}ssland.
\newblock {T}he {P}rofessionalization of the {I}ntolerable: {P}opular {C}rime
  {J}ournalism in {N}orway.
\newblock {\em Journalism Studies}, 8(1):137--152, 2007.

\bibitem{Roumani22}
Y.~Roumani.
\newblock {D}etection {T}ime of {D}ata {B}reaches.
\newblock {\em Computers \& Security}, 112:102508, 2022.

\bibitem{Ruohonen22IS}
J.~Ruohonen and K.~Hjerppe.
\newblock {T}he {GDPR} {E}nforcement {F}ines at {G}lance.
\newblock {\em Information Systems}, 106:101876, 2022.

\bibitem{Ruohonen24PRR}
J.~Ruohonen, K.~Hjerppe, and K.~Kortesuo.
\newblock {C}risis {C}ommunication in the {F}ace of {D}ata {B}reaches.
\newblock {A}rchived manuscript. Available online in June:
  \url{https://arxiv.org/abs/2406.01744}, 2024.

\bibitem{Sacco95}
V.~F. Sacco.
\newblock {M}edia {C}onstruction of {C}rime.
\newblock {\em Annals of the American Academy of Political and Social Science},
  539:141--154, 1995.

\bibitem{Schlackl22}
F.~Schlackl, N.~Link, and H.~Hoehle.
\newblock {A}ntecedents and {C}onsequences of {D}ata {B}reaches: {A}
  {S}ystematic {R}eview.
\newblock {\em Information \& Management}, 59(4):103638, 2022.

\bibitem{Sinanaj14}
G.~Sinanaj.
\newblock {N}ews {M}edia {S}entiment of {D}ata {B}reaches.
\newblock In {\em Proceedings of the 20th Americas Conference on Information
  Systems (AMCIS 2014)}, Savannah, 2014. AIS.
\newblock {A}vailable online in June 2024:
  \url{https://core.ac.uk/download/pdf/301362123.pdf}.

\bibitem{Smolej08}
M.~Smolej and J.~Kivivuori.
\newblock {C}rime {N}ews {T}rends in {F}inland:{A} {R}eview of {R}ecent
  {R}esearch.
\newblock {\em Journal of Scandinavian Studies in Criminology and Crime
  Prevention}, 9(2):202--219, 2008.

\bibitem{Spanos16}
G.~Spanos and L.~Angelis.
\newblock {T}he {I}mpact of {I}nformation {S}ecurity {E}vents to the {S}tock
  {M}arket: {A} {S}ystematic {L}iterature {R}eview.
\newblock {\em Computers \& Security}, 58:216--229, 2016.

\bibitem{Syed19}
R.~Syed.
\newblock {E}nterprise {R}eputation {T}hreats on {S}ocial {M}edia: {A} {C}ase
  of {D}ata {B}reach {F}raming.
\newblock {\em Journal of Strategic Information Systems}, 28:257--274, 2020.

\bibitem{NLTK24}
{T}he {N}atural {L}anguage~{T}oolkit (NLTK).
\newblock {V}ersion 3.4.5.
\newblock Available online in September: \url{http://www.nltk.org}, 2024.

\bibitem{Thomas22}
L.~Thomas, I.~Gondal, T.~Oseni, and S.~S. Firmin.
\newblock {A} {F}ramework for {D}ata {P}rivacy and {S}ecurity {A}ccountability
  in {D}ata {B}reach {C}ommunications.
\newblock {\em Computers \& Security}, 116:102657, 2022.

\bibitem{Trussler14}
M.~Trussler and S.~Soroka.
\newblock {C}onsumer {D}emand for {C}ynical and {N}egative {N}ews {F}rames.
\newblock {\em The International Journal of Press/Politics}, 19(3):360--379,
  2014.

\bibitem{vanKrieken22}
K.~{van Krieken}.
\newblock {S}tory {C}haracter, {N}ews {S}ource or {V}ox {P}op?
  {R}epresentations and {R}oles of {C}itizen {P}erspectives in {C}rime {N}ews.
\newblock {\em Journalism}, 39(9):1975--1994, 2022.

\bibitem{Velasquez20}
D.~Vel\'asquez, S.~Medina, G.~Yamada, P.~Lavado, M.~{Nunez-del-Prado},
  H.~{Alatrista-Salas}, and J.~Morz\'an.
\newblock {I} {R}ead the {N}ews {T}oday, {O}h {B}oy: {T}he {E}ffect of {C}rime
  {N}ews {C}overage on {C}rime {P}erception.
\newblock {\em World Development}, 136:105111, 2020.

\bibitem{YanChua20}
N.~W.~J. Yan and H.~N. Chua.
\newblock {A} {P}ath {A}nalysis {M}odel to {I}dentify the {E}ffects of {S}ocial
  {M}edia, {N}ews {M}edia and {D}ata {B}reach on {D}ata {P}rotection
  {R}egulation {A}wareness.
\newblock In {\em Proceedings of the 2nd International Conference on Artificial
  Intelligence in Engineering and Technology (IICAIE 2020)}, pages 1--6, Kota
  Kinabalu, 2020. IEEE.

\end{thebibliography}

\end{document}